# Polaritonic coherent perfect absorption based on strong critical coupling between quasi-bound state in the continuum and exciton

Xin Gu,[1,2] Xing Liu,[1,2] Xiaofei Yan,[1,2] Qi Lin,[1,3] Wenjuan Du,[1] Lingling Wang[3] and Guidong Liu[1,3, *]

[1]School of Physics and Optoelectronics, Xiangtan University, Xiangtan 411105, China

[2]Hunan Engineering Laboratory for Microelectronics, Optoelectronics and System on a Chip, Xiangtan University, Xiangtan 411105, China

[3]School of Physics and Electronics, Hunan University, Changsha 410082, China

*gdliu@xtu.edu.cn

**ABSTRACT:** Enhancement of light-matter interactions is of great importance for many nanophotonic devices, and one way to achieve it is to feed energy perfectly to the strongly coupled system. Here, we propose gap-perturbed dimerized gratings based on bulk $WS_2$ for flexible control of the strong coupling between quasi-bound state in the continuum (quasi-BIC) and exciton. The simulation results show that when a gap perturbation is introduced into the system resulting in the Brillouin zone folding, BIC transforms into quasi-BIC whose quality factor ($Q$-factor) is related to the value of gap perturbation. The strong coupling results in the anti-crossover behavior of the absorption spectra, and thus a Rabi splitting energy of 0.235 eV is obtained. Temporal coupled-mode theory is employed to analyze the strong coupling system, and the absorption of the system could be enhanced, by modulating the damping rate of quasi-BIC to make the system satisfy the strong critical coupling condition. Furthermore, polaritonic coherent perfect absorption is achieved by using the two-port source excitation. This work could provide ideas for enhancing light-matter interactions and a promising prospect for polariton-based light-emitting or lasing devices.

## I. INTRODUCTION

Exciton-polariton [1-2] is currently a fascinating topic and also a significant direction for studying strong coupling [3-4]. A "strong coupling" state is achieved when the system interacts at a rate faster than the dissipation of light and emitter, which leads to the formation of half-light, half-matter quasiparticles called polaritons. The resulting mixed semi-light and semi-matter states combine with the advantage of photons and exciton to achieve strong interparticle interactions and generate some interesting physical phenomena such as Bose-Einstein condensates [5], superfluidity [6], polariton lasers [7], and solitons [8]. In recent years, the strong coupling between the exciton in transition-metal dichalcogenides (TMDCs) with the trapped light have been widely researched because the considerable binding energy of TMDCs make it possible to explore the light-matter interaction at room temperature [3, 9, 10]. For further enhancing the light-matter interaction, many efforts are attempted to reduce the mode volume or achieve high-$Q$ mode because the splitting energy is closely related to the ratio of $Q$-factor and mode volume $V$ [11]. The plasmonic mode beyond diffraction limit could greatly suppress the mode volume and manipulate the light in sub-wavelength scale. A number of hybrid systems are proposed to investigate the plasmon-exciton coupling in varied TMDCs materials with metallic nanostructure [12-15]. Nevertheless, a plasmonic mode suffers from the intrinsic metallic loss and optical radiative loss, so it is difficult to get particularly high $Q$-factor, thus limiting their splitting energy.

Photonic bound state in the continuum (BIC) has attracted great attention owing to infinite value of $Q$-factor and strong electric field enhancement. BIC is generally lying in the continuum radiation spectrum, but does not leak into free space. A real BIC is regard as a mathematical object, which is difficult to be observed in the spectrum, and fortunately, in practice, it can be implemented as a quasi-BIC [16, 17]. The quasi-BIC can be supported by all dielectric metasurface or optical micro-cavity, which can not only attain high $Q$-factor, but also provide strong electric enhancement. Hence, the quasi-BIC is a promising candidate to achieve the strong exciton-polariton coupling. To date, some related researches have achieved strong coupling between quasi-BIC and exciton in monolayer TMDCs [18-20]. These works have made considerable progress in the field of studying the strong coupling between quasi-BIC and TMDC exciton, which achieved the maximum Rabi splitting energy of 65 meV in monolayer WS$_2$ heterostructures.

Maximizing absorption is significant in many systems and has many applications, such as photoluminescence enhancement [21], absorber [22], and laser [23]. The absorption of a system is usually positively correlated with the field enhancement, which can be provided by the quasi-BIC [24]. As the TMDCs exciton couple to the quasi-BIC, the strong electric field confinement provided by the quasi-BIC can greatly enhance the absorption of the hybrid system [25]. And beyond that, the absorption could be further enhanced to reach a maximum value of 0.5 for a single port input, by optimizing the structural parameters to make the system satisfy the critical coupling conditions [26, 27]. The essential condition for the realization of critical coupling is the matching of damping losses [28], or the balance between radiation losses and the non-radiation loss rate of the hybrid system [29]. Many previous studies have shown that the damping rate of quasi-BIC has a wide range of control through appropriate methods or means, such as changing the structural symmetry [17]. Another way to achieve ultra-high absorption, that is the coherent light incident by two beams can be completely absorbed in the dielectric resonator with a certain loss [30, 31]. According to the mutual phase of the incident beams, the system can also achieve complete absorption, which is called coherent perfect absorption, and the optical absorption is further enhanced.

In this work, we propose a polaritonic system based on bulk WS$_2$ for flexible control of the strong coupling between the quasi-BIC and exciton. As a gap perturbation is introduced into the system resulting in the Brillouin zone folding, BIC transforms into high $Q$-factor quasi-BIC. The spectral overlap and spatial superposition between the quasi-BIC and exciton can be realized, resulting in a typical anti-crossing behavior in the absorption spectrum and a significant Rabi splitting of 0.235 eV. To satisfy the strong critical coupling condition, we employ temporal coupled mode theory (TCMT), enabling the damping rate of the quasi-BIC matches that of exciton by modulating the gap, thereby enhancing the absorption of the system. Furthermore, to further enhance the absorption of the system, polaritonic coherent perfect absorption (CPA) is explored under two-port source excitation.

## II. STRUCTURE AND PRINCIPLE

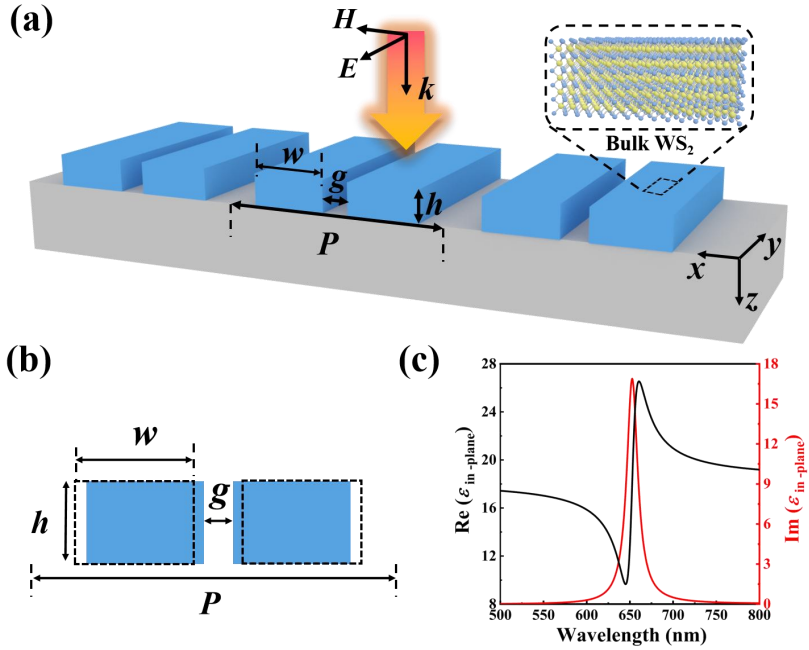

FIG. 1. (a) Schematic diagram of dimerized gratings composed of bulk $WS_2$. (b) The *xz* cross-sectional of unit cell of the structure, with the dashed boxes corresponding to the case *g* = 145 nm. (c) Real and imaginary parts of the in-plane dielectric function ($\varepsilon_{in\text{-}plane}$) of bulk $WS_2$.

As the sketch map shows in Fig. 1(a), we propose dimerized gratings composed of the bulk $WS_2$, and consider a homogenous background with permittivity 1 for simplicity. The inset illustrates the atomic layer structure of bulk $WS_2$. The period is *P*, the width and height of two grating fingers are *w* and *h*, and the air gap between two grating fingers is *g*. The structural parameters are set as follows: *P* = 500 nm, *w* = 105 nm, *h* = 61 nm and *g* = 145 nm. The *xz* cross-sectional view of the structure is shown in Fig. 1(b). The proposed dimerized gratings can be considered as two identical and unperturbed gratings with period *P*/2. When a gap perturbation is introduced, for instance, each other grating finger laterally shifts the same distance in the opposite direction, gap-perturbed dimerized gratings with period *P* are formed which are comprised of two identical gratings with period *P*/2. The finite-difference time-domain (FDTD) Solutions is used to investigate the electromagnetic characteristics of the proposed structure. In the simulation, a normal incident plane wave polarized along the *y*-direction is used. A periodic boundary condition in the *x*-direction and perfectly matched layers (PMLs) condition along the *z*-direction are embraced.

The material of choice is the anisotropic bulk $WS_2$, and its in-plane permittivity is described as the follow formula [32, 33]:

$$\varepsilon_{\text{in-plane}} = \varepsilon_E + f_0 \frac{\omega_{\text{ex}}^2}{\omega_{\text{ex}}^2 - \omega^2 - i\gamma_{\text{ex}}\omega}, \qquad (1)$$

where $\varepsilon_E = 18$ is the background permittivity, $f_0$ represents the oscillator strength, and $\omega_{\text{ex}} = 1.9$ eV and $\gamma_{\text{ex}} = 0.045$ eV are the resonance wavelengths of the exciton and full width, respectively. The out-of-plane permittivity is $\varepsilon_{\text{out-of-plane}} = 7$. Assuming $f_0 = 0$, it indicates that exciton is not introduced and the material in the plane can be considered as a dielectric with a permittivity of 18. When $f_0 = 0.4$, the exciton of bulk $WS_2$ can be considered. The real and imaginary parts of the in-plane permittivity of bulk $WS_2$ calculated according to Eq. (1) are shown in Fig. 1(c). The imaginary part presents a sudden trend near 652 nm (1.9 eV), which makes the bulk $WS_2$ have a large line width at 1.9 eV, so it can be used as an ideal strong coupling material.

## III. RESULTS AND DISCUSSIONS
### A. Optical properties of the quasi-BIC

To gain the physics of BIC, the band structure of dimerized gratings with $g = 145$ nm is calculated, as shown in Fig. 2(a). It reveals that such a grating support a BIC mode at point Γ of the first Brillouin zone. The inset in Fig. 2(a) illustrates the normalized electric field distribution of the eigenmode at 652 nm, and the black dashed boxes represent the structure. As a gap perturbation is introduced, two identical gratings with period $P/2$ are conversion into gap-perturbed dimerized gratings with period $P$, thereby making the BIC transform to quasi-BIC. The $Q$-factor of gap-perturbed dimerized gratings as a function of gap $g$ is shown in Fig. 2(b). It can be observed that BIC transforms into quasi-BIC whose $Q$-factor grows with the increase or decrease of gap. The physical mechanism can be explained as the doubling of the period due to the change of the gap, resulting in the Brillouin zone folding, which finally makes the previously bound state to be excitable by free-space illumination [34]. In addition, Figs. 2(c) and (d) show the cross sections of the electric field distributions of BIC and quasi-BIC, where $g$ is 145 nm and 127 nm, respectively. Since there is no radiation channel, the field of the BIC is completely localized inside the proposed structure. However, only a tiny perturbation of the quasi-BIC relative to the BIC also results in a significant radiation of the system to the air, along with a huge decrease in the maximum field enhancement. It is worth noting that the dipole cloud of FDTD Solutions is used as the excitation source in this part.

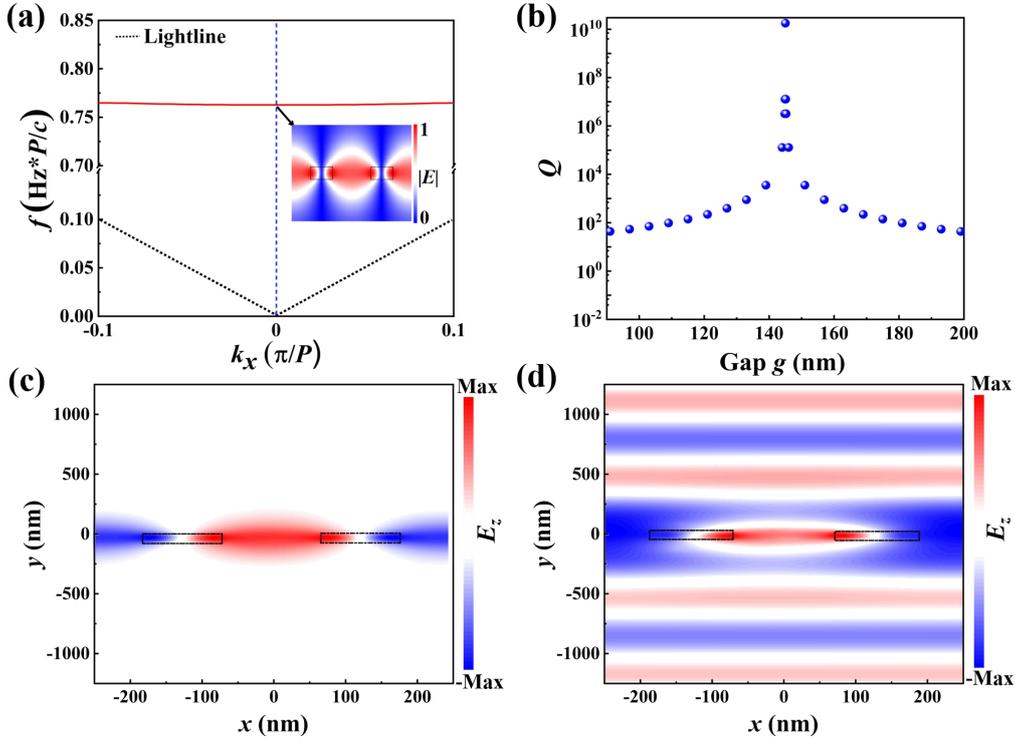

FIG. 2. (a) Band structure of the system associated with BIC. Inset: Normalized field distribution of BIC mode, where the black boxes indicate the structure. (b) The $Q$-factor of quasi-BIC as a function of gap $g$. The electric field ($E_z$) of (c) BIC ($g$ = 145 nm) and (d) quasi-BIC ($g$ = 127 nm).

## B. Realization of strong coupling and analysis based on temporal coupled-mode theory

In order to achieve the strong coupling of the quasi-BIC and exciton, the structural parameters are optimized to excite a high-$Q$ quasi-BIC at the wavelength of the exciton (652 nm). In Fig. 3(a), the transmission spectrum of the gap-perturbed dimerized gratings with $g$ = 127 nm is calculated, which exhibits a sharp asymmetric line shape and the resonance wavelength is 652 nm. It should be pointed out that the structural parameters which can achieve wavelength matching are not unique. The effects of height $h$ and period $P$ on the transmission spectra of gap-perturbed dimerized gratings are studied and the results are shown in Figs. 3(b) and 3(c), respectively. It can be obviously found that the quasi-BIC exhibit a red-shift while its linewidth has almost no change, as the height $h$ increases from 50 to 70 nm. For the change caused by the period, the quasi-BIC shows a relatively small red shift and the line width gradually magnifies when the period $P$ increases from 580 nm to 720 nm, as shown in Fig. 3(c).

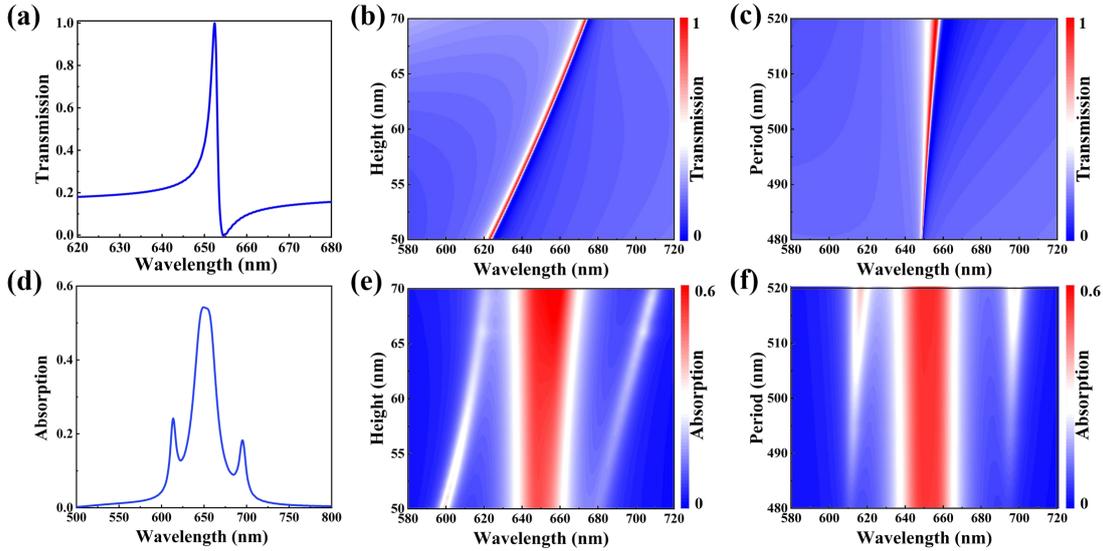

FIG. 3. (a) The transmission spectrum of the quasi-BIC with $g$ = 127 nm. Transmission spectra of the quasi-BIC with respect to (b) height and (c) period. (d) Absorption spectrum of hybrid mode with gap $g$ = 127 nm and height $h$ = 61 nm. Absorption spectra of hybrid modes at different (e) height and (f) period.

To prove the strong coupling between quasi-BIC and exciton in bulk $WS_2$, the material parameter $f_0$ = 0.4 material in Eq. (1) are considered. The absorption spectrum of the coupled system is shown in Fig. 3(d), where the absorption is defined by equation $A = 1 - R - T$. As one can see, there are three peaks in the absorption spectrum, and the wavelengths are 614 nm, 652 nm, 695 nm, respectively. The new peaks appearing at 614 nm and 695 nm are induced by the strong coupling of quasi-BIC and exciton, and a Rabi splitting energy reaches approximately 0.235 eV. Quasi-BIC typically have large local field enhancement, and the quasi-BIC supported by our proposed structure has a maximum electric field enhancement in bulk $WS_2$, which enables exciton to fully contact and coherently coupling with it. This is the reason why the splitting energy obtained by our proposed structure is much greater than that of some previous studies on monolayer $WS_2$ [18-20, 25]. In addition, the significant absorption peak at 652 nm between the two new branches can be mainly attributed to exciton. The mapping of the absorption spectrum for an increase in height from 50 nm to 70 nm is calculated and shown in Fig. 3(e). One can find that the wavelength of the intermediate peak almost does not change with the height, while it can be seen from Fig. 3(b) that the quasi-BIC is redshifted with the increase of height. Therefore, it can be concluded that the absorption peak in the middle is mainly contributed by exciton. It can also be observed from Fig. 3(e) that the anti-crossing lines occur. With the height $h$ increases, the hybrid modes are red-shifted, so that the splitting energy values are changed accordingly. The effect of period on the strong coupling is also studied, as shown in Fig. 3 (f), the value of splitting energy change little, while the absorption of the hybrid modes gradually increases as period increase from 480 nm to 520 nm.

Temporal coupled-mode theory could be used to describe the system. The system has a coupling constant $d$ and is externally excited as a resonator. $|A_Q|$ and $|A_E|$ represent the amplitude of the quasi-BIC and exciton, respectively. And $\omega_Q$ ($\omega_E$), $\gamma_Q$ ($\gamma_E$), and $\Omega$ are the corresponding resonance frequency and damping rate, and the coupling coefficient. The input wave amplitudes $|s^+\rangle = (s_1^+, s_2^+)$ and the corresponding output wave amplitudes $|s_1^-\rangle, |s_2^-\rangle$ can be obtained by the equations [25, 36]

$$\frac{dA_E}{dt} = (i\omega_E - \gamma_E)A_E + i\Omega A_Q, \quad (2)$$

$$\frac{dA_Q}{dt} = (i\omega_Q - \gamma_Q)A_Q + i\Omega A_E + \left(\langle d|^*\right)|s^+\rangle, \quad (3)$$

$$|s^-\rangle = C|s^+\rangle + A_Q|d\rangle, \tag{4}$$

Where the input and output waves are linked as $|s^-\rangle = S(\omega)|s^+\rangle$. Combining the above formulas, one can obtain

$$S(\omega) = C - \frac{i(\omega - \omega_Q) + \gamma_E}{(\omega - \omega_+)(\omega - \omega_-)} D, \tag{5}$$

Where the matrices $C$ and $D = |d\rangle(\langle d|)^*$ can be obtained from Ref. [35]. The eigenvalue of hybrid states is

$$E_{UB, LB} = \hbar\omega_\pm = \frac{\hbar}{2}[\omega_E + \omega_Q + i(\gamma_Q + \gamma_E)]$$
$$\pm \frac{\hbar}{2}\sqrt{4\Omega^2 + [(\omega_E - \omega_Q) + i(\gamma_E - \gamma_Q)]^2}, \tag{6}$$

where $E_{UB, LB}$ denotes the upper branch (UB) and lower branch (LB) of energy. Under zero detuning condition, that is, when $\omega_Q = \omega_E = \omega_0$, the Eq. (6) can be simplified as

$$E_{UB,LB} = \hbar\omega_0 + \frac{\hbar}{2}[i(\gamma_Q + \gamma_E)] \pm \frac{\hbar}{2}\sqrt{4\Omega^2 - (\gamma_E - \gamma_Q)^2}, \tag{7}$$

The Rabi splitting energy is defined by the energy separation at the anti-crossing point, which describe as $\hbar\Omega_R = \hbar\sqrt{4\Omega^2 - (\gamma_Q - \gamma_E)^2}$. The numerical simulation results in Fig. 3(d) show that when $h$ = 61 nm, the wavelength of quasi-BIC is equal to the exciton, $\omega_Q = \omega_E = \omega_0 = 652$ nm, and the Rabi splitting energy is $\hbar\Omega_R = 0.235$ eV. The damping rate of exciton is $\gamma_E = 0.04$ eV, while that of quasi-BIC can be obtained from Fig. 3(a) with $\gamma_Q = 0.003$ eV, then one can obtain $\hbar\Omega = 0.118$ eV. Here, $\gamma_E = (\omega_1 - \omega_2)/2$, $\omega_1$, $\omega_2$ represent the frequency which the half maximum absorption of exciton occurs. And $\gamma_Q = (\omega_p - \omega_d)/2$, $\omega_p$, $\omega_d$ are the frequencies corresponding to the peak and dip of the transmission spectrum, respectively.

The blue asterisk in Fig. 4(a) represents the theoretical calculation results of Eq. (6), which is basically consistent with the simulation result in Fig. 3(e). Besides, the theoretical results correspond to the numerical results of the orange solid line. When Rabi splitting is greater than the dissipative energy of the hybrid system, $\hbar\Omega_R > \hbar(\gamma_Q + \gamma_E)$, strong coupling and anti-crossover behavior are achieved.

Hence, the criteria of strong coupling are defined as

$$\hbar\Omega > \frac{\hbar}{2}|\gamma_Q - \gamma_E|, \quad \hbar\Omega > \hbar\sqrt{\frac{1}{2}(\gamma_Q^2 + \gamma_E^2)}, \tag{8}$$

Besides, the quasi-BIC and exciton fraction weights in the upper and lower branches are obtained, using equality [20]

$$|\alpha|^2 = \frac{1}{2}(1 \pm \frac{\delta}{\sqrt{\delta^2 + 4\Omega^2}}), \tag{9}$$

$$|\beta|^2 = \frac{1}{2}(1 \mp \frac{\delta}{\sqrt{\delta^2 + 4\Omega^2}}), \tag{10}$$

where the detuning $\delta = \hbar\omega_Q - \hbar\omega_E$. As shown in Fig. 4(b), in the upper and lower branches, the fraction weights of quasi-BIC and exciton change in different trends. For example, the weight of quasi-BIC increases with the height of the upper branch, indicating that more photons are coupled.

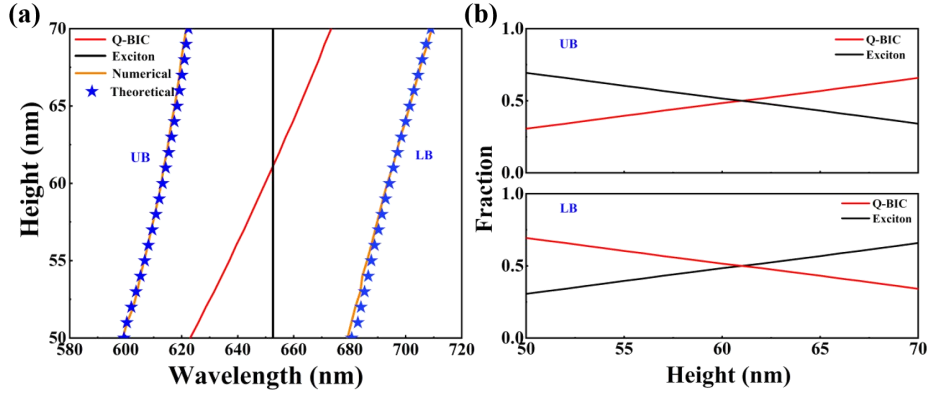

FIG. 4. (a) Theoretical and numerical results of upper branch (UB) and lower branch (LB) of the strong coupling. (b) The fraction curves of exciton (black solid line) and quasi-BIC (red solid line), respectively.

## C. Strong critical coupling and coherent perfect absorption

When the S-matrix satisfying its determinant equal to 0, the system can be regard as an ideal interferometer. Consider $\omega_Q = \omega_E = \omega_0$, by calculating det $S(\omega) = 0$, one can obtain

$$\bar{E}_{UB,LB} = \hbar\omega_0 + \frac{\hbar}{2}[i(-\gamma_Q + \gamma_E)] \pm \frac{\hbar}{2}\sqrt{4\Omega^2 - (\gamma_E + \gamma_Q)^2}. \tag{11}$$

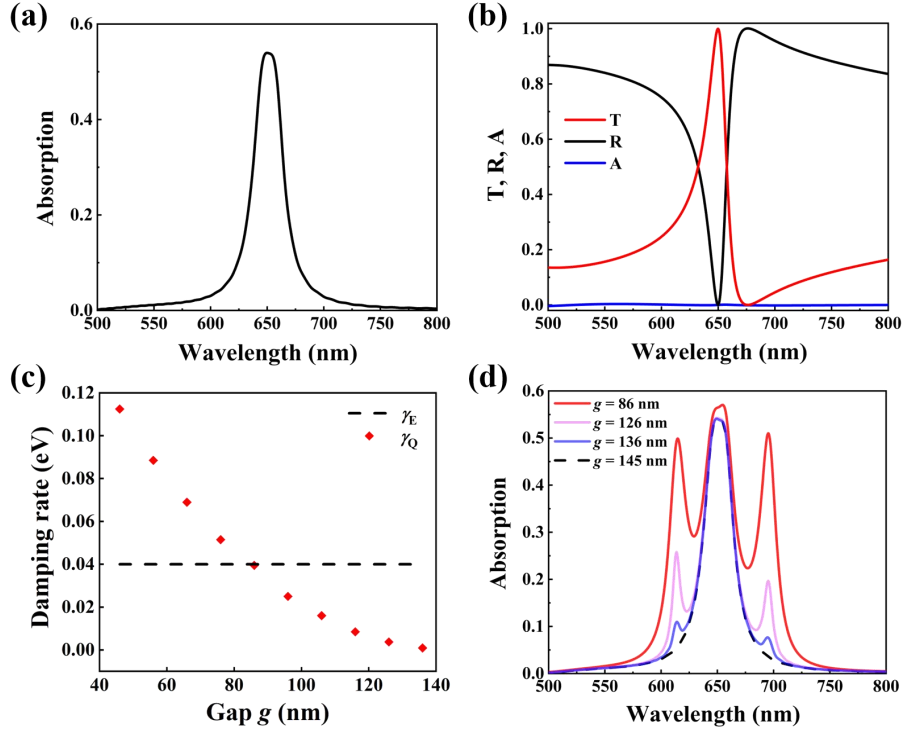

FIG. 5. (a) Absorption spectrum of the grating with $g$ = 145 nm. (b) The transmission, reflection and absorption spectra of the quasi-BIC under $g$ = 86 nm. (c) The damping rate $\gamma_Q$ and $\gamma_E$ as the functions of $g$. (d) The absorption spectra of the gratings with different $g$.

Theoretically, under the condition $\gamma_E = \gamma_Q$, the system will achieve strong critical coupling [36]. To satisfy this condition, we keep height $h$ = 61 nm and investigate the dependence of the damping rate of quasi-BIC on the gap $g$. We calculate the absorption spectrum of bulk $WS_2$ grating with $g$ = 145 nm so as not to introduce quasi-BIC in Fig. 5(a), thus the damping rate of exciton $\gamma_E$ = 0.04 eV can be obtained. Figure 5(b) indicates the spectrum of a quasi-BIC without considering the exciton, and its linewidth is adjusted to match that of exciton. The damping rate of exciton and quasi-BIC are shown in Fig. 5(c) as a function of gap, it is obvious that $\gamma_E = \gamma_Q$ = 0.04 eV when $g$ = 86 nm. In order to verify whether the system satisfies strong critical coupling condition, the absorption spectra with varied $g$ are calculated, as shown in Fig. 5(d). The results reveal that the absorption peaks of hybrid modes are significantly enhanced. The maximum absorption of the system reaches 0.5, proves the existence of strong critical coupling.

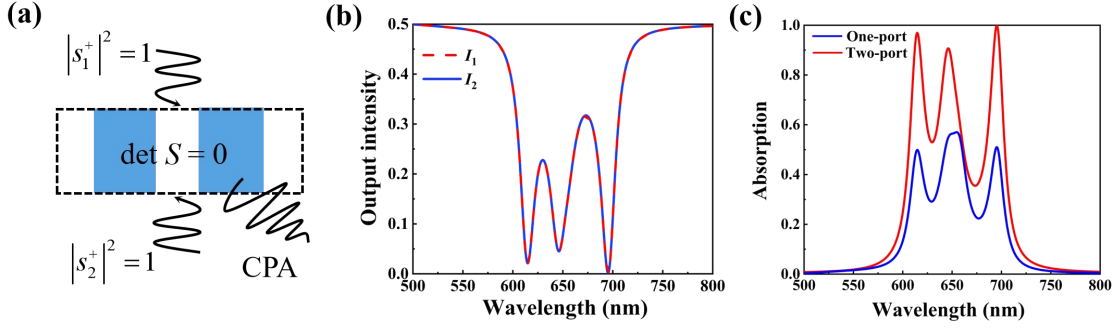

FIG. 6. (a) Schematic diagram of coherent perfect absorption. (b) The output intensities $I_1$ and $I_2$ of the system under two-port source excitation. (c) The absorption spectra of the system under one-port and two-port source excitation, respectively.

To further enhance the absorption of the hybrid system, the polaritonic CPA is explored based on the strong critical coupling state shown in Fig. 6(a). The sum of output intensity is $|s_1^-|^2 + |s_2^-|^2 \approx 0$ at the wavelengths of 614.7 and 695.2 nm by choosing the two-port source phase difference π with input intensities $|s_1^+|^2 = |s_2^+|^2 = 1$. The output intensities $I_1$ and $I_2$ are obtained by placing two monitors behind the source. Simulation results in Fig. 6(b) show that the output intensities at resonant wavelengths almost reach 0, indicating the achievement of polaritonic CPA. It enables the maximum absorption of the system is equal to 1, as shown in Fig. 6(c).

## IV. CONCLUSIONS

In conclusion, we propose the gap-perturbed dimerized gratings based on bulk $WS_2$ to investigate the strong coupling between the quasi-BIC and exciton. As a gap perturbation is introduced into the system resulting in the Brillouin zone folding, BIC transforms into quasi-BIC whose Q-factor is related to the value of gap perturbation. The anti-crossover behavior appears in the absorption spectrum and a relatively large Rabi splitting energy about 0.235 eV is achieved. Combined the FDTD Solutions and TCMT, the strong coupling is analyzed numerically and theoretically. The absorption of the system could be greatly enhanced, by modulating the damping rate of quasi-BIC equal to that of exciton to make the system satisfy the strong critical coupling condition. Polaritonic CPA is also achieved and the maximum absorption of the system reaches 1. Our results will potentially pave the way for the realization of novel TMDCs polaritonic devices and high-performance absorber.

## ACKNOWLEDGEMENTS

This work is supported by the National Natural Science Foundation of China (62205278, 11947062,62105276) and Hunan Provincial Natural Science Foundation of China (2020JJ5551, 2021JJ40523, 2020JJ5550).